# A Practical Method for Automated Modeling and Parametric Stability Analysis of VSC with Periodical Steady State

Chen Zhang, Jon Are Suul, Marta Molinas

***Abstract*—Linear Time Periodic (LTP) framework-based analysis of Voltage Source Converters (VSCs) is becoming popular, a driven factor is that many of the existing VSC applications inevitably exhibit the periodic steady-state (PSS), e.g., VSCs with unbalanced grid connections and the operation of a modular multilevel converter (MMC). In these studies, acquisition of the VSC's PSS conditions is a necessary precondition for proper linearization and stability analysis, and the efficiency of this process is particularly important for parametric studies. To this end, this work develops a computational method for automating the LTP analyses of VSCs with an integrated PSS solver. The core of the method lies in a unified *frequency-domain iteration* process that is developed by applying the generalized averaging principle. Given by this, modeling, stability analysis, as well as the solution of PSS conditions can be unified in one process. Algorithmic implementation of the method in MATLAB is elaborated. Application of the obtained tool in impedance generation and parametric stability test is presented with an unbalanced grid-tied VSC as the exemplification. Finally, PSCAD/EMTDC simulations further consolidate the validity of the results.**

## SYMBOLS AND NOMENCLATURE

$x(t)$: real-valued variable

$\dot{x}(t)$: time derivative of a variable

$\boldsymbol{x}$: real-valued vector

$X_{(k)}$: $k$-th Fourier coefficient of $x(t)$

$\boldsymbol{X}_{(k)}$: $k$-th spectral vector of $\boldsymbol{x}$

$\boldsymbol{\mathcal{X}}$: vector collection of $\boldsymbol{X}_{(k)}$

$\boldsymbol{\mathcal{X}}(k)$: $k$-th element of vector $\boldsymbol{\mathcal{X}}$

$\boldsymbol{A}(t)$: time-domain matrix

$\boldsymbol{\mathcal{A}}$: Toeplitz formatted matrix of $\boldsymbol{A}(t)$

$\vec{\boldsymbol{u}}_{\alpha\beta}$: space vector or complex vector, e.g., as $\vec{\boldsymbol{u}}_{\alpha\beta} = u_\alpha + \mathrm{j}u_\beta$

$\alpha\beta$: $\alpha\beta$ reference frame

*abc*: *abc* reference frame

*dq*: *dq* reference frame

## I. INTRODUCTION

Recently, field experiences from the operation of large-scale wind farms [1] and photovoltaic power plants [2] have shown that Voltage Source Converters (VSCs) are prone to trigger oscillation issues when connected to weak AC grids.

To systematically study such issues, small-signal stability assessments of the grid-tied VSC using either the linearized state-space model or the impedance model has been widely discussed. In detail, the state-space modelling is commonly used as a basis for modal analysis of oscillations in power systems [3]. From the modal analysis, oscillation characteristics and the contribution of states to specific oscillation modes can be analyzed via the participation factor (PF) technique, resulting in a powerful tool for oscillation analysis [4], [5]. The impedance-based method, on the other hand, is a frequency domain approach that is gaining popularity in oscillation analysis recently mainly because: 1) many of the occurred oscillation issues in practice are caused by device interactions in circuits, such port-interaction-based stability issues can be more directly analyzed by modeling the VSC and other devices as impedances in circuits; 2) moreover, impedance analysis can be fulfilled either with analytic models or the measured frequency responses, which is more flexible to practical applications. Through years of endeavor, impedance modeling techniques for a three-phase two-level VSC are almost well established, and various impedance models with different insights into the VSC have been proposed in this regard [6]-[11]. With the impedance models, stability of a single grid-tied VSC can be equally analyzed by a "source-load system" using the Nyquist criterion [12], [13]. However, generalization of this analysis from the single converter case to an impedance network with multi converters should be cautious, where some applicability issues such as the selection of partition point, the observability of branch oscillations [14], etc., may arise. Such issues are largely related with improper circuit operations and aggregations of the impedance network. Thus, to alleviate these issues, the multi-source-load method can be considered [15].

The above works are in general developed within the linear time invariant (LTI) framework, where a crucial premise is that

C. Zhang is with the Department of Electrical Engineering, Shanghai Jiao Tong University (SJTU), Shanghai, 200240, China (email: nealbc@sjtu.edu.cn).
J. A. Suul and M. Molinas are with the Department of Engineering Cybernetics, Norwegian University of Science and Technology (NTNU), 7034 Trondheim, Norway, where J. A. Suul is also with the SINTEF Energy Research, 7491 Trondheim, Norway (email: Jon.A.Suul@sintef.no, marta.molinas@ntnu.no).

the modeled system can be represented by constant-valued steady-state (CSS) variables. However, continuously increasing utilization of power electronics is accelerating the need for modelling and analysis of systems including conversion units or operating conditions that cannot be easily modelled in a time-invariant framework, e.g., the single-phase converter and the modular multilevel converter (MMC), of which the systems inevitably exhibit the periodic steady-state (PSS).

To address the need for modeling and analysis of PSS systems, the linear-time periodic (LTP) theory [16], [17] is often applied. In this respect, a variety of modeling works have been conducted for different types of converters, e.g., the thyristor-based converter [18], the single-phase VSC [19]-[21], three-phase VSC with unbalances [22], [23], as well as the MMC [24], [25], etc. According to these works, the general process of the LTP-based analysis can be summarized as three main steps: 1) establish a continuous and differentiable state-space model of the system using the switching average principle; 2) linearize the averaged model upon the assumingly obtained PSS conditions; 3) extract the system's PSS conditions and insert them into the linearized model for LTP analysis. From this procedure it can be observed that the extraction of PSS conditions is an essential process for LTP analysis, and this is usually fulfilled by either lengthy time-domain simulations (e.g., [19]) or analytical calculations with simplifying assumptions (e.g., [20], [26]). However, the simulation-based PSS extraction is not suitable for parametric studies, not only due to the lack of efficiency (as the PSS conditions need to be frequently updated according to the parameter changes) but also for not being able to extract unstable PSS conditions, while the analytic approximation-based approach involves tedious calculation and is not readily generalizable to other cases.

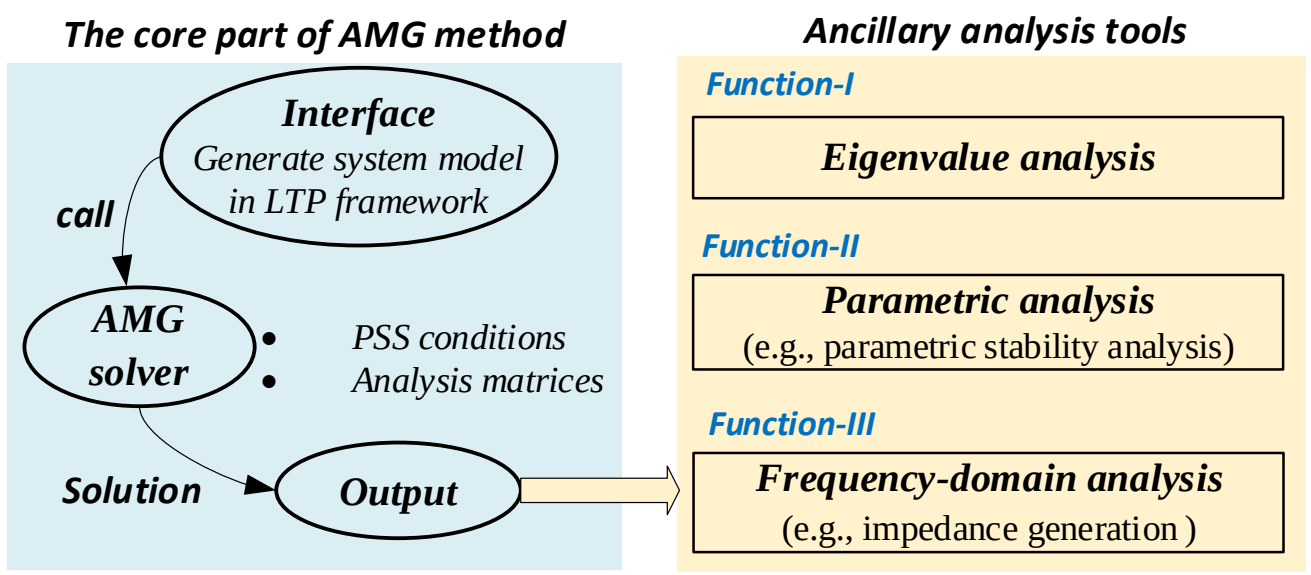

Fig. 1 Overall structure of the AMG method and the resulting AMG tools

In fact, the general challenge of PSS extraction can be addressed by numerical methods using iterative techniques [28], [29]. In the spirit of this, this paper will present a practical computation framework for automated modeling and analysis of PSS-based converter systems, which results in a tool of this work, referred to as the automatic model generation (AMG). The overall structure of the tool to be developed is shown in Fig. 1, for brevity, it is referred to as the automatic model generation (AMG) method. As indicated by the figure, the presented AMG tool can directly generate the small-signal models necessary for eigenvalue-based or impedance-based analysis using the results of the PSS solver. To elaborate on the development and application of the tool, the following sections are arranged:

Section II introduces the AMG method and its algorithmic implementation. Demonstration of the functions in Fig. 1 is shown in Section III using a simplified grid-VSC system; while the application of the tool in a more detailed grid-VSC system is presented in Section IV. Finally, Section VI draws the main conclusions of this paper.

## II. Formulation of the AMG Method and its Algorithmic Implementation

In this section, the basis of the LTP theory necessary for the AMG development will be introduced first, then the formulation of the AMG method will be elaborated.

### A. Overview of the LTP method

For a generic nonlinear system

$$\begin{cases} \dot{\boldsymbol{x}} = \boldsymbol{f}(t, \boldsymbol{x}, \boldsymbol{u}) \\ \boldsymbol{y} = \boldsymbol{g}(t, \boldsymbol{x}, \boldsymbol{u}) \end{cases} \tag{1}$$

excited by a $T$-periodical input vector $\boldsymbol{u}$, its steady-state will be $T$-periodic, i.e., the presence of PSS. Linearizing (1) will typically result in an LTP system written as

$$\begin{cases} \Delta\dot{\boldsymbol{x}} = \boldsymbol{A}(t)\Delta\boldsymbol{x} + \boldsymbol{B}(t)\Delta\boldsymbol{u} \\ \Delta\boldsymbol{y} = \boldsymbol{C}(t)\Delta\boldsymbol{x} + \boldsymbol{D}(t)\Delta\boldsymbol{u} \end{cases} \tag{2}$$

where $\boldsymbol{A}(t), \boldsymbol{B}(t), \boldsymbol{C}(t), \boldsymbol{D}(t)$ are $T$-periodic matrices. Analysis of (2) can be performed in either time- or frequency-domain.

*1) The time-domain method:* Floquet theory indicates that for an LTP system, there exists a linear transformation $x(t) = R(t)z(t)$ such that the transformed system $\dot{z}(t) = Q \cdot z(t)$ is LTI, where $Q = A(t)R(t) - \dot{R}(t)$. In this way, stability of the original LTP system can be equally analyzed by its LTI equivalence. However, finding such a transformation $R(t)$ via analytic methods is not a trivial issue, where the calculation of $Q$ matrix mostly relies on numeric methods [30].

*2) The frequency-domain method*: Hill provides a frequency-domain approach for LTP system analysis [16]. The core idea is to use the exponentially-modulated-periodic (EMP) signal (e.g., $x(t) = e^{st}\sum_m X_m e^{jm\omega_1 t}$ ) to reformulate (2) into a so-called harmonic state-space (HSS) model:

$$\begin{cases} (s\mathcal{I} + \mathcal{N}_{\mathrm{blk}}) \cdot \mathcal{X} = \mathcal{A} \cdot \mathcal{X} + \mathcal{B} \cdot \mathcal{U} \\ \mathcal{Y} = \mathcal{C} \cdot \mathcal{X} + \mathcal{D} \cdot \mathcal{U} \end{cases} \tag{3}$$

$\mathcal{N}_{\mathrm{blk}} = diag(-\mathrm{j}N\omega_1\boldsymbol{I}, ..., \mathrm{j}\omega_1\boldsymbol{I}, ..., +\mathrm{j}N\omega_1\boldsymbol{I})$ is a block-diagonal matrix, where the dimension of the identity matrix $\boldsymbol{I}$ equals the dimension of the state-vector $\boldsymbol{x}(t)$ ; $\mathcal{A}, \mathcal{B}, \mathcal{C}, \mathcal{D}, \mathcal{I}$ are the Toeplitz formatted matrices of $\boldsymbol{A}(t), \boldsymbol{B}(t), \boldsymbol{C}(t), \boldsymbol{D}(t)$ , and $\boldsymbol{I}$ ; $\mathcal{X} = \left[\boldsymbol{X}_{(-N)}, ..., \boldsymbol{X}_{(0)}, ..., \boldsymbol{X}_{(+N)}\right]$ is the collection of vectors of $\boldsymbol{X}_{(k)}$ , where $\boldsymbol{X}_{(k)}$ denotes the $k$-th spectral vector of $\boldsymbol{x}(t)$ ; the same definition applies to $\mathcal{U}$ and $\mathcal{Y}$ . With the HSS model, stability of (2) can be better analyzed by calculating the eigenvalues of $\mathcal{A} - \mathcal{N}_{\mathrm{blk}}$ [20], which is similar to the analysis of an LTI system.

It should be noted that both of the methods rely on the knowledge of matrix $\boldsymbol{A}(t)$ which is a function of the system's PSS conditions. In this work, fast and efficient acquisition of the PSS conditions will be achieved by the AMG solver, which will be shown next.

## *B. Method for the AMG solver*

The AMG solver is a unified frequency-domain iteration process that can be efficiently solved by Newton's method. Derivation of its formulation is in general based on the generalized averaging (GA) technique [31], the process will be shortly given below.

First, (1) at *k*-th harmonic can be written as

$$\frac{d\langle \boldsymbol{x}\rangle_k}{dt} = -\mathrm{j}k\omega_1\langle \boldsymbol{x}\rangle_k + \langle \boldsymbol{f}(t,\boldsymbol{x},\boldsymbol{u})\rangle_k \tag{4}$$

where the linear operator $\langle\cdot\rangle_k$ denotes the averaging against the *k*-th harmonic, e.g., $e^{\mathrm{j}k\omega_1 t}$, in which $\omega_1$ is the fundamental frequency of the system. The resulting $\langle \boldsymbol{x}\rangle_k$ then denotes the *k*-th time-varying Fourier coefficient of $\boldsymbol{x}(t)$. See the derivation of (4) from (1) in Appendix. A. From (4) it can be seen that the PSS conditions of (1) are now represented by their Fourier coefficients, which are time-invariant constants in steady-state and thus can be found by Newton's method. Development of (4) towards the final iteration model is shown next.

The *i*-th step iteration-model of (4) can be written as:

$$\boldsymbol{0} = -\mathrm{j}k\omega_1\langle \boldsymbol{x}\rangle_k^{(i)} - \mathrm{j}k\omega_1\Delta\langle \boldsymbol{x}\rangle_k^{(i)} + \langle \boldsymbol{f}(t,\boldsymbol{x},\boldsymbol{u})\rangle_k^{(i)} + \left\langle \frac{\partial \boldsymbol{f}(t,\boldsymbol{x},\boldsymbol{u})}{\partial \boldsymbol{x}}\cdot\Delta\boldsymbol{x}\right\rangle_k^{(i)} \tag{5}$$

For PSS calculation, (5) usually represents a closed-loop system, thus the change of the overall system's input is omitted. Further, the partial derivative $\frac{\partial \boldsymbol{f}(t,\boldsymbol{x},\boldsymbol{u})}{\partial \boldsymbol{x}} = \boldsymbol{A}(t) = \sum_{m=-N}^{N} A_m e^{\mathrm{j}m\cdot\omega_1 t}$ can be obtained by the Fourier series of $\boldsymbol{A}(t)$. Similarly, the Fourier series of $\Delta\boldsymbol{x}$ can be written as $\Delta\boldsymbol{x} = \sum_{k=-N}^{N}\Delta\langle \boldsymbol{x}\rangle_k e^{\mathrm{j}k\cdot\omega_1 t}$.

Then, insert these equations into the last term of (5), yields:

$$\boldsymbol{0} = -\mathrm{j}k\omega_1\langle \boldsymbol{x}\rangle_k^{(i)} - \mathrm{j}k\omega_1\Delta\langle \boldsymbol{x}\rangle_k^{(i)} + \langle \boldsymbol{f}(t,\boldsymbol{x},\boldsymbol{u})\rangle_k^{(i)} + \sum_{m=-N}^{N} A_{k-m}^{(i)}\Delta\langle \boldsymbol{x}\rangle_m^{(i)} \tag{6}$$

when deriving the above equation, the following equation obtained from the GA is applied:

$$\left\langle \sum_{k=-N}^{N}\left(\sum_{k=-N}^{N} A_{k-m}\Delta\langle \boldsymbol{x}\rangle_m\right)e^{\mathrm{j}k\cdot\omega_1 t}\right\rangle_k^{(i)} = \sum_{m=-N}^{N} A_{k-m}^{(i)}\Delta\langle \boldsymbol{x}\rangle_m^{(i)} \tag{7}$$

Since (6) should be solved for each considered harmonic, i.e., within the range $k\in[-N,N]$, thus a matrix representation of the *i*-th step iteration model can be obtained as (8). And its compact version can be written below:

$$\begin{cases}\boldsymbol{0} = \left(\boldsymbol{\mathcal{A}}^{(i)} - \boldsymbol{\mathcal{N}}_{\mathrm{blk}}\right)\Delta\boldsymbol{\mathcal{X}}^{(i)} + \boldsymbol{\mathcal{F}}^{(i)} - \boldsymbol{\mathcal{N}}_{\mathrm{blk}}\boldsymbol{\mathcal{X}}^{(i)} \\ \boldsymbol{\mathcal{X}}^{(i+1)} = \Delta\boldsymbol{\mathcal{X}}^{(i)} + \boldsymbol{\mathcal{X}}^{(i)}\end{cases} \tag{9}$$

This is the final frequency-domain iteration model of the AMG solver. Its algorithm will be introduced next.

## *C. Generic algorithm for the AMG solver*

When solving (9) iteratively, $\boldsymbol{\mathcal{A}}^{(i)}$ and $\boldsymbol{\mathcal{F}}^{(i)}$ should be updated from the results of the previous iteration. This process involves the time- to frequency-domain transformations, for which the following steps take place internally:

**Step 1**: Simulate $A(t)$ and $\boldsymbol{f}\left(\boldsymbol{x}^{(i)},\boldsymbol{u}^{(i)}\right)$ over a short duration ($T_{\mathrm{AMG}}$) with a time-step ($h_{\mathrm{AMG}}$) using $\boldsymbol{x}^{(i)}$, $\boldsymbol{u}^{(i)}$.

**Step 2:** Transform $A(t)$ and $\boldsymbol{f}\left(\boldsymbol{x}^{(i)},\boldsymbol{u}^{(i)}\right)$ into frequency-domain, obtaining $\boldsymbol{\mathcal{A}}^{(i)}$ and $\boldsymbol{\mathcal{F}}^{(i)}$.

**Step 3:** Solve (9) for $\boldsymbol{\mathcal{X}}^{(i+1)}$. Check if the tolerance is reached, e.g., $\left\|\Delta\boldsymbol{\mathcal{X}}^{(i+1)}\right\|\le\varepsilon$, Otherwise, goes to **Step 4.**

**Step 4:** Acquire the new time-domain state-vector $\boldsymbol{x}^{(i+1)}$ from $\boldsymbol{\mathcal{X}}^{(i+1)}$ by the inverse Fourier transform and go to **Step 1.**

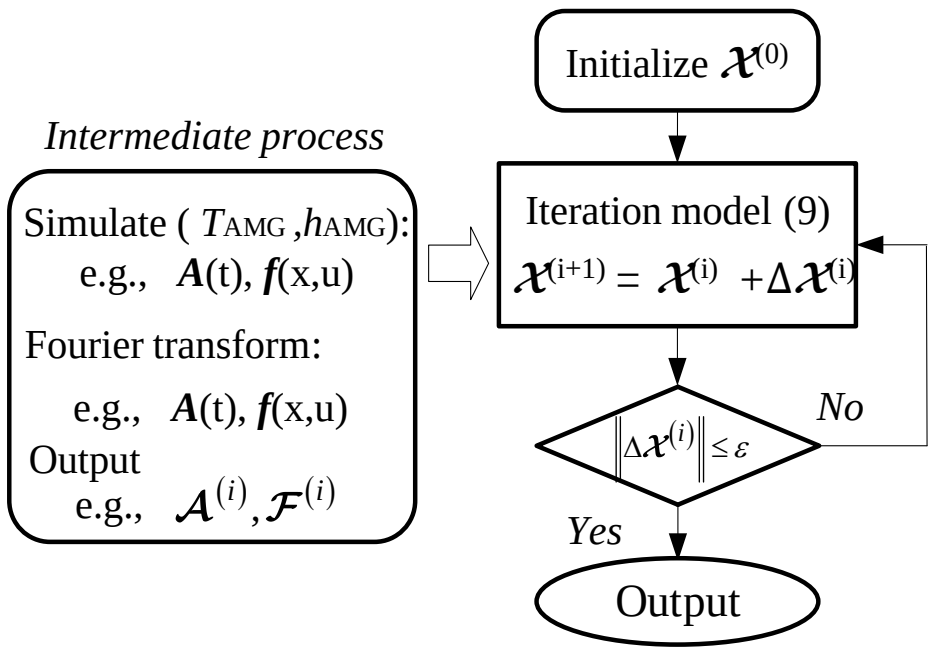

Fig. 2 Flowchart of the generic algorithm for the AMG solver

**Discussion on the outputs of the AMG solver:** Outputs if the AMG solver will not only contain the PSS conditions (i.e., $\boldsymbol{x}^{(i)}$ and its Fourier coefficiencies $\boldsymbol{\mathcal{X}}^{(i)}$ ) but also include

$$-\boldsymbol{\mathcal{N}}_{\mathrm{blk}}\underbrace{\begin{bmatrix}\Delta\langle \boldsymbol{x}\rangle_{-N}^{(i)}\\ \vdots \\ \Delta\langle \boldsymbol{x}\rangle_{0}^{(i)} \\ \vdots \\ \Delta\langle \boldsymbol{x}\rangle_{N}^{(i)}\end{bmatrix}}_{\Delta\boldsymbol{\mathcal{X}}^{(i)}} - \boldsymbol{\mathcal{N}}_{\mathrm{blk}}\underbrace{\begin{bmatrix}\langle \boldsymbol{x}\rangle_{-N}^{(i)}\\ \vdots \\ \langle \boldsymbol{x}\rangle_{0}^{(i)} \\ \vdots \\ \langle \boldsymbol{x}\rangle_{N}^{(i)}\end{bmatrix}}_{\boldsymbol{\mathcal{X}}^{(i)}} + \underbrace{\begin{bmatrix}\langle \boldsymbol{f}(t,\boldsymbol{x},\boldsymbol{u})\rangle_{-N}^{(i)}\\ \vdots \\ \langle \boldsymbol{f}(t,\boldsymbol{x},\boldsymbol{u})\rangle_{0}^{(i)} \\ \vdots \\ \langle \boldsymbol{f}(t,\boldsymbol{x},\boldsymbol{u})\rangle_{N}^{(i)}\end{bmatrix}}_{\boldsymbol{\mathcal{F}}^{(i)}} + \underbrace{\begin{bmatrix}A_0 & A_{-1} & & \cdots & A_{-2N}\\ A_{+1} & \ddots & \ddots & & \vdots \\ & \ddots & A_0 & \ddots & \\ \vdots & & \ddots & \ddots & A_{-1} \\ A_{2N} & \cdots & & A_{+1} & A_0\end{bmatrix}}_{\boldsymbol{\mathcal{A}}^{(i)}}\underbrace{\begin{bmatrix}\Delta\langle \boldsymbol{x}\rangle_{-N}^{(i)}\\ \vdots \\ \Delta\langle \boldsymbol{x}\rangle_{0}^{(i)} \\ \vdots \\ \Delta\langle \boldsymbol{x}\rangle_{N}^{(i)}\end{bmatrix}}_{\Delta\boldsymbol{\mathcal{X}}^{(i)}} = \boldsymbol{0} \tag{8}$$

relevant by-products (e.g., $\mathcal{A}^{(i)},\mathcal{B}^{(i)},\mathcal{C}^{(i)},\mathcal{D}^{(i)}$) for various LTP analyses as mentioned in Fig. 1. For example, the stability can be analyzed by calculating the eigenvalues of $\mathcal{A}^{(i)}-\mathcal{N}_{\text{blk}}$, while the input-output frequency response can be generated by using all the matrices $\mathcal{A}^{(i)},\mathcal{B}^{(i)},\mathcal{C}^{(i)},\mathcal{D}^{(i)}$. These functions will be demonstrated later using case studies.

Finally, a flowchart summarizing the above algorithm of the AMG solver is shown in Fig. 2, and the corresponding configuration used in this paper is given in Table I.

TABLE I CONFIGURATION FOR THE AMG SOLVER

| | |
|---|---|
| Harmonic-order in (9): $N = 4$ | Simulation step: $h_{\text{AMG}} = 50\ \mu$s |
| Sim-duration: $T_{\text{AMG}} = 0.02$ s | Tolerance $\varepsilon = 0.001$ |
| Note: Initial values for iteration are set according to reference values of controls while the remaining values are filled with zeros | |

**Remarks on the configuration**: In most cases, $T_{\text{AMG}} = T$ is adequate for the iteration, where $T$ is the period of the system, e.g., $T_{\text{AMG}} = 20$ *ms* corresponds to a 50 Hz system. Since the iteration is based on a switching-averaged model, a larger $h_{\text{AMG}}$ can be used compared to EMT simulations. To ensure the speed of convergence as well as accuracy, this paper applies $h_{\text{AMG}}$ = 50 $\mu$ s. Next, $N$ denotes the highest order of harmonics considered in the iteration model. In principle, the higher the order the better the model converges to its time-domain equivalence [32]. However, this will drastically increase the model complexity and may not be necessary. In this work, $N$ = 4 is selected based on the following considerations: 1) the iteration model is switching averaged, thus the system is free from switching and sideband harmonics; 2) the converter systems typically involve ac/dc signal modulations, thus to better capture this process $N$ = 3 is suggested (e.g., see a single-phase VSC case in [21] and an MMC case in [25]); 3) to alleviate potential numeric issues arising from the matrix truncation at the tails, one more order is reserved as the margin, which leads to the adoption $N$ = 4 of this work.

## D. Interface for the AMG solver

When introducing the AMG solver, it is assumed that models, e.g., $\boldsymbol{A}(t)$ and $\boldsymbol{f}(\boldsymbol{x},\boldsymbol{u})$ are known ahead. Usually, they can be obtained by analytic derivations. However, when the system's order becomes large, this process will become tedious and cumbersome, e.g., when deriving $\boldsymbol{A}(t)$, a lengthy linearization process is required. To address this issue and promote the applicability of the AMG method, the interface module as shown in Fig. 1 is assigned and intended for achieving a system-wide linearization. In this regard, software with the capability of symbolic calculation will be of great benefit. In this next, MATLAB will be used serving as an exemplification for showing how the interface module is implemented and integrated with the AMG solver.

## E. Implementation of the AMG method in MATLAB

### 1) Symbolic modeling

First, formulate a generic nonlinear system (1) using the command "*syms*". Then, perform a system-wide linearization on (1) using the command "*jacobian*", as a result, the LTP model (2) is obtained, i.e., the matrices $\boldsymbol{A}(t),\boldsymbol{B}(t),\boldsymbol{C}(t),\boldsymbol{D}(t)$.

### 2) Symbolic to numeric conversion

Transform the symbolic functions into parametric functions using the command "*matlabFunction*". This will increase the efficiency in numerically communicating with the AMG solver.

### 3) Call the AMG solver

This process follows the generic algorithm in Fig. 2, however, when it comes to the Fourier analysis of e.g., $\boldsymbol{A}(t)$ and $\boldsymbol{f}(\boldsymbol{x},\boldsymbol{u})$, the MATLAB command "*FFT*" can be used.

Procedures of 1) and 2) constitute the "Interface" module in Fig. 1, while 1)-3) leads to the AMG tool that will be used in this paper. Besides, following a similar procedure, the AMG method can also be implemented in other software.

# III. DEMONSTRATION OF AMG METHOD AND ITS FUNCTIONALITIES USING A SIMPLIFIED GRID-VSC SYSTEM

First, a simplified grid-VSC system (i.e., "Case system I" in Fig. 3) will be analyzed to demonstrate the application of the AMG tool and the resulting functionalities. To this end, the following three types of asymmetries are considered for the excitation of PSS conditions in this section:

1) Asymmetry in the three-phase filter;
2) Asymmetry in the three-phase grid inductance;
3) Asymmetry in the three-phase grid voltage.

In which, the first two conditions are imposed by using the following relations: $L_{fa}=L_{fb}$, $L_{fc}=k_{\text{sym_c}}L_{fa}$, and $L_{ga}=L_{gb}$, $L_{gc}=k_{\text{sym_g}}L_{ga}$, where $k_{\text{sym_c}}$ and $k_{\text{sym_g}}$ are defined symmetry factors; while the third condition is imposed by changing the magnitude/phase of grid voltages, i.e., $u_{\text{ga}}$, $u_{\text{gb}}$, $u_{\text{gc}}$.

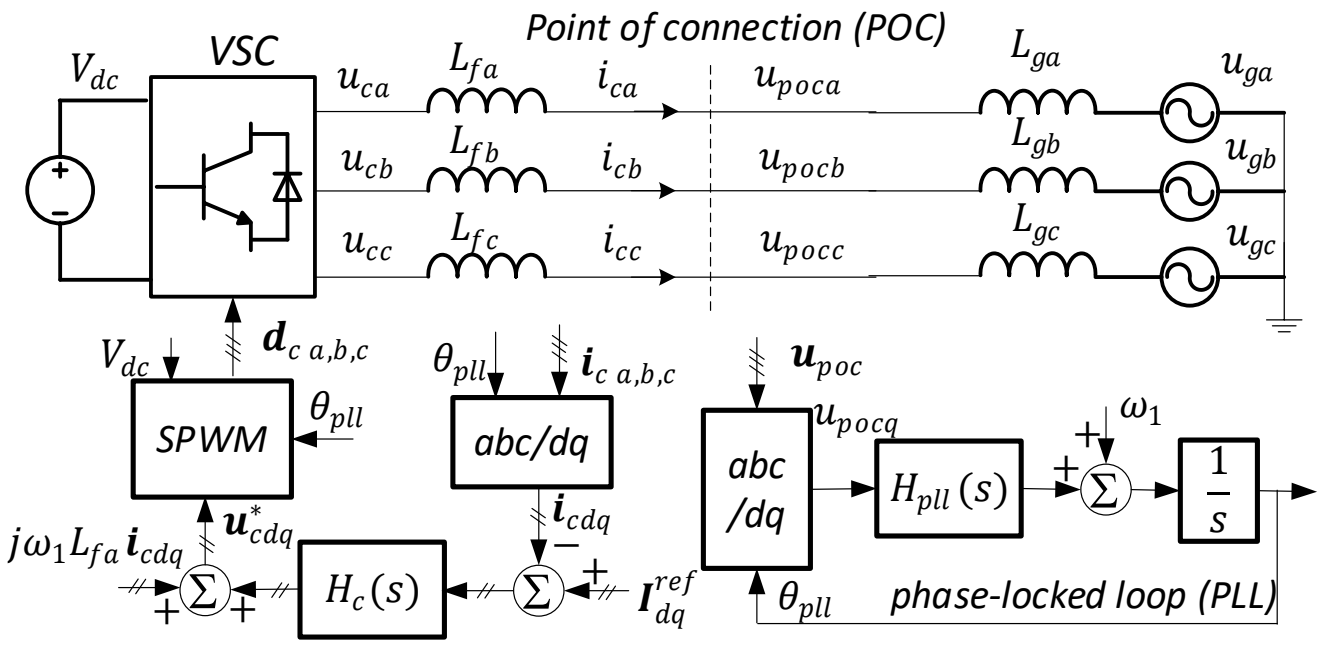

Fig. 3 Case system I: a simplified grid-tied VSC system ($H_c = k_{pc} + k_{ic}/s$, $H_{pll} = k_{ppll} + k_{ipll}/s$)

## A. Vector-based state-space modeling

Since the system's state-space model is the input of the AMG tool, it should be derived in advance. When modeling a three-phase system, it is common practice to use the space-vector as a compact representation. More often, the space-vector can be written in the $\alpha\beta$-frame as $\vec{\boldsymbol{u}}_{\alpha\beta}=u_{\alpha}+\mathrm{j}u_{\beta}$, where the zero-axis is exempted from analysis as there is no path for the zero-axis current to flow in this case. Furthermore, when the three-phase is *not* symmetric, the conjugate of $\vec{\boldsymbol{u}}_{\alpha\beta}$ (i.e., $\vec{\boldsymbol{u}}^*_{\alpha\beta}$ ) will coexistent, which should be considered in the modeling. It is worth mentioning that the state-space modeling work can be performed in other frames (e.g., *abc*- or *dq*-frame), which will not affect the applicability of the AMG method.

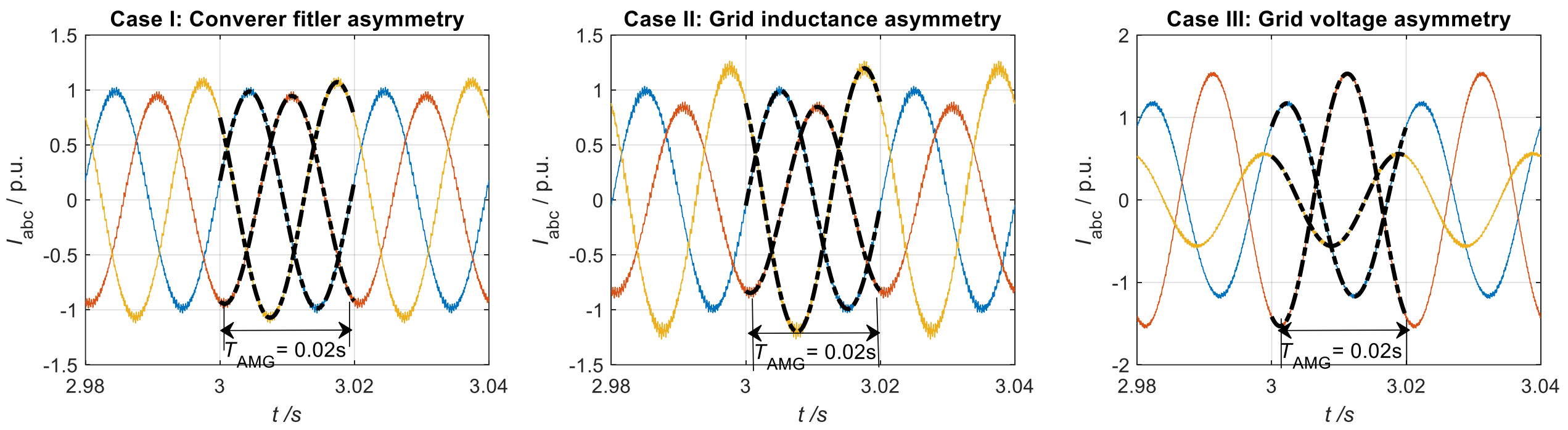

Fig. 4  Test of PSS extraction under the Case system I with different types of three-phase asymmetries (solid line: PSCAD, dotted line: AMG method)

Based on the above illustration, the vector-based state-space model of the Case system I will be given as follows.

*1) PLL model*

$$\begin{cases} \dot{x}_{\text{pll}} = k_{\text{ipll}} u_{\text{pocq}} \\ \dot{\delta}_{\text{pll}} = k_{\text{ppll}} u_{\text{pocq}} + x_{\text{pll}} \end{cases} \tag{10}$$

where $x_{\text{pll}}$ is the integral state of $H_{\text{pll}}(s)$ and $\delta_{\text{pll}} = \theta_{\text{pll}} - \omega_1 t$. In (10), the *q*-axis voltage can be written as

$$u_{\text{pocq}} = \frac{1}{2\text{j}}\left( e^{-\text{j}(\omega_1 t+\delta_{\text{pll}})} \vec{u}_{\text{poc_}\alpha\beta} - e^{\text{j}(\omega_1 t+\delta_{\text{pll}})} \vec{u}^*_{\text{poc_}\alpha\beta} \right) \tag{11}$$

*2) Current control model*

Applying the vector representation, the current control system can be modeled as (12) and (13), where (12) is the state equation of CC and (12) is the converter output voltage.

$$\begin{cases} \dot{\vec{x}}_{\text{cdq}} = k_{\text{ic}}\left( \vec{I}^{ref}_{\text{dq}} - e^{-\text{j}(\omega_1 t+\delta_{\text{pll}})} \vec{i}_{\text{c}\alpha\beta} \right) \\ \dot{\vec{x}}^*_{\text{cdq}} = k_{\text{ic}}\left( \vec{I}^{ref*}_{\text{dq}} - e^{-\text{j}(\omega_1 t+\delta_{\text{pll}})} \vec{i}^*_{\text{c}\alpha\beta} \right) \end{cases} \tag{12}$$

$$\begin{cases} \vec{u}_{\text{c}\alpha\beta} = k_{\text{pc}}\left( \vec{I}^{ref}_{\text{dq}} e^{\text{j}(\omega_1 t+\delta_{\text{pll}})} - \vec{i}_{\text{c}\alpha\beta} \right) + \vec{x}_{\text{cdq}} e^{\text{j}(\omega_1 t+\delta_{\text{pll}})} + \text{j}\omega_1 L_{fa} \vec{i}_{\text{c}\alpha\beta} \\ \vec{u}^*_{\text{c}\alpha\beta} = k_{\text{pc}}\left( \vec{I}^{ref*}_{\text{dq}} e^{-\text{j}(\omega_1 t+\delta_{\text{pll}})} - \vec{i}^*_{\text{c}\alpha\beta} \right) + \vec{x}^*_{\text{cdq}} e^{-\text{j}(\omega_1 t+\delta_{\text{pll}})} - \text{j}\omega_1 L_{fa} \vec{i}^*_{\text{c}\alpha\beta} \end{cases} \tag{13}$$

*3) Asymmetrical filter and grid models*

Vector-based filter and grid models can be transformed from their three-phase models (i.e., in the *abc*-frame). Take the filter model as an example, its vector-based model can be derived as

$$\boldsymbol{L}_{\text{f},\text{C}_{\alpha\beta}} = \boldsymbol{T}_{\alpha\beta/\text{C}_{\alpha\beta}} \boldsymbol{T}_{\text{abc}/\alpha\beta}\, diag\left( L_{\text{fa}}, L_{\text{fb}}, k_{\text{sym_c}} L_{\text{fa}} \right) \boldsymbol{T}_{\alpha\beta/\text{abc}} \boldsymbol{T}_{\text{C}_{\alpha\beta}/\alpha\beta} \tag{14}$$

where all the applied transformation matrices are given in Appendix B. Based on this, the vector-based state-space model of the grid inductance and the filter can be obtained as:

$$\boldsymbol{L}_{\text{f},\text{C}_{\alpha\beta}} \begin{bmatrix} \dot{\vec{i}}_{\text{c}\alpha\beta} \\ \dot{\vec{i}}^*_{\text{c}\alpha\beta} \end{bmatrix} = \begin{bmatrix} \vec{u}_{\text{c}\alpha\beta} - \vec{u}_{\text{poc_}\alpha\beta} \\ \vec{u}^*_{\text{c}\alpha\beta} - \vec{u}^*_{\text{poc_}\alpha\beta} \end{bmatrix}, \boldsymbol{L}_{\text{g},\text{C}_{\alpha\beta}} \begin{bmatrix} \dot{\vec{i}}_{\text{g}\alpha\beta} \\ \dot{\vec{i}}^*_{\text{g}\alpha\beta} \end{bmatrix} = \begin{bmatrix} \vec{u}_{\text{poc_}\alpha\beta} - \vec{u}_{\text{g}\alpha\beta} \\ \vec{u}^*_{\text{poc_}\alpha\beta} - \vec{u}^*_{\text{g}\alpha\beta} \end{bmatrix} \tag{15}$$

*4) Formulation of the closed- and open-loop models*

Assembling (10)-(15), the closed-loop model of the grid-VSC can be formulated as:

$$\begin{cases} \dot{\boldsymbol{x}}_{\text{cl}} = \boldsymbol{f}_{\text{cl}}\left(t, \boldsymbol{x}_{\text{cl}}, \boldsymbol{u}_{\text{g}}\right) \\ \boldsymbol{y}_{\text{cl}} = \boldsymbol{g}_{\text{cl}}\left(\boldsymbol{x}_{\text{cl}}\right) \end{cases} \tag{16}$$

where $\boldsymbol{x}_{\text{cl}} = \left[ \vec{i}_{\text{c}\alpha\beta}, \vec{i}^*_{\text{c}\alpha\beta}, \vec{x}_{\text{cdq}}, \vec{x}^*_{\text{cdq}}, \delta_{\text{pll}}, x_{\text{pll}} \right]^T$ , $\boldsymbol{y}_{\text{cl}} = \left[ \vec{i}_{\text{c}\alpha\beta}, \vec{i}^*_{\text{c}\alpha\beta} \right]^T$ , $\boldsymbol{u}_{\text{g}} = \left[ \vec{u}_{\text{g}\alpha\beta}, \vec{u}^*_{\text{g}\alpha\beta} \right]^T$ . Besides, the open-loop model of VSC can also be obtained by considering (10)-(15), which yields:

$$\begin{cases} \dot{\boldsymbol{x}}_{\text{vsc}} = \boldsymbol{f}_{\text{vsc}}\left(t, \boldsymbol{x}_{\text{vsc}}, \boldsymbol{u}_{\text{poc}}\right) \\ \boldsymbol{y}_{\text{vsc}} = \boldsymbol{g}_{\text{vsc}}\left(\boldsymbol{x}_{\text{vsc}}\right) \end{cases} \tag{17}$$

where $\boldsymbol{x}_{\text{vsc}}$ are identical to $\boldsymbol{x}_{\text{cl}}$ because for this case there are no shunt components between the VSC and the grid.

Then, (16) and (17) can be implemented in the AMG tool according to the procedures given in Section II.E.

Next, the AMG-based PSS extraction and ancillary analysis tools (i.e., the Function I-III in Fig. 1) will be demonstrated, where the applied AMG configuration and the main system parameters are shown in Table I and Table II.

## *B. Test of PSS extraction*

To achieve a comprehensive evaluation, the following severely asymmetric cases will be analyzed:

**Case I**:  VSC filter asymmetry (i.e., $k_{sym_c} = 0.1$). In which the condition $L_{fa} = L_{fb}, L_{fc} = 0.1\, L_{fa}$ is imposed.

**Case II**:  Grid inductance asymmetry (i.e., $k_{sym_g} = 0.1$). In which the condition: $L_{ga} = L_{gb}, L_{gc} = 0.1\, L_{ga}$ is imposed.

**Case III**: Grid voltage unbalance (i.e., $|U_{g\beta}| = 0.5\, p.u.$). Based on the condition: $U_{g\alpha} = 1\angle 0^{\circ} p.u.$ and $U_{g\beta} = 0.5\angle{-90^{\circ}} p.u.$ used in AMG tool, the following three-phase grid voltage for simulation can be obtained: $U_{ga} = 1\angle 0^{\circ} p.u.$, $U_{gb} = 0.66\angle{-139.1^{\circ}} p.u.,$ *and* $U_{gc} = 0.66\angle 139.1^{\circ} p.u.$

A comparison of current waveforms obtained from the AMG tool and the PSCAD/EMTDC simulations under the above-introduced asymmetric conditions is shown in Fig. 4. From the results, it can be clearly seen that the extracted PSS conditions from the AMG method is consistent with the simulation, moreover, the AMG only requires a *T*-period simulation in the iteration (i.e., mainly for evaluating $A(t)$ and $\boldsymbol{f}\left(\boldsymbol{x}^{(i)}, \boldsymbol{u}^{(i)}\right)$), thus it is in principle more efficient than numeric simulations, which usually takes even longer time for reaching the steady-state when the system is poorly damped. Furthermore, it is worthy noting again that, time-domain simulations cannot extract unstable PSS conditions, limiting its application scope compared to the AMG method.

TABLE II DEFAULT SYSTEM PARAMETERS

| | |
|---|---|
| Circuit parameters | |
| Rating: $S_N$ = 2 MVA | VSC inductance $L_{fa,b,c} = 7.58 \times 10^{-5}$ H |
| Voltage: $U_N = 0.690$ kV | Grid inductance $L_{ga,b,c} = 1.89 \times 10^{-4}$ H |
| Filter capacitor (Case system II) $C_{a,b,c} = 500 \times 10^{-6}$ F | Resistor (Case system II) $R_{ca,b,c} = 0.01$ ohm |
| Control parameters | |
| $I_{dq}^{ref} = (1 + 0j)$ p.u. | CC BD: $\alpha_c = 200$ Hz |
| Switching frequency: $f_{sw} = 6000$ Hz | PQ BD: $\alpha_s = 20$ Hz (Case system II) |
| Base values for per unit system conversion | |
| $U_{base}$ = 0.563 kV | $I_{base}$ = 2.368 kA |
| $S_{base}$ = 2 MVA | $L_{base}$ = 0.758 mH |
| $f_{base}$ = 50 Hz | $\omega_{base} = 2\pi f_{base}$ rad/s |

Note: 1) $L_{gc}$ , $L_{fc}$ can be varied by manipulating $k_{sym_g}$ , $k_{sym_c}$.
2) Bandwidth (BD) in relation to PI gains are given in Appendix-C.

## C. *Test of Function I: Eigenvalue analysis*

### 1) *Eigenvalue calculation*

As mentioned earlier, the outputs of the AMG solver contain useful by-products (e.g., $\mathcal{A}_{cl}^{(i)}, \mathcal{B}_{cl}^{(i)}, \mathcal{C}_{cl}^{(i)}, \mathcal{D}_{cl}^{(i)}$ ) that can be utilized for a variety of LTP analyses. For instance, the stability analysis can be fulfilled by calculating the eigenvalues of $\mathcal{A}_{cl}^{(i)} - \mathcal{N}_{blk}$ [20], and an example of this is shown in Fig. 5. It can be seen that the eigenvalues contain the frequency-shifted copies, which is a trait of the LTP system. However, for stability assessment, the real part of the weakest mode (i.e., $Re[\lambda_{weakest}]$) is of utmost importance. Specifically, $Re[\lambda_{weakest}] > 0$ denotes an *unstable* system, otherwise, the system is *stable*. Hence, $Re[\lambda_{weakest}]$ is used for later parametric stability assessment.

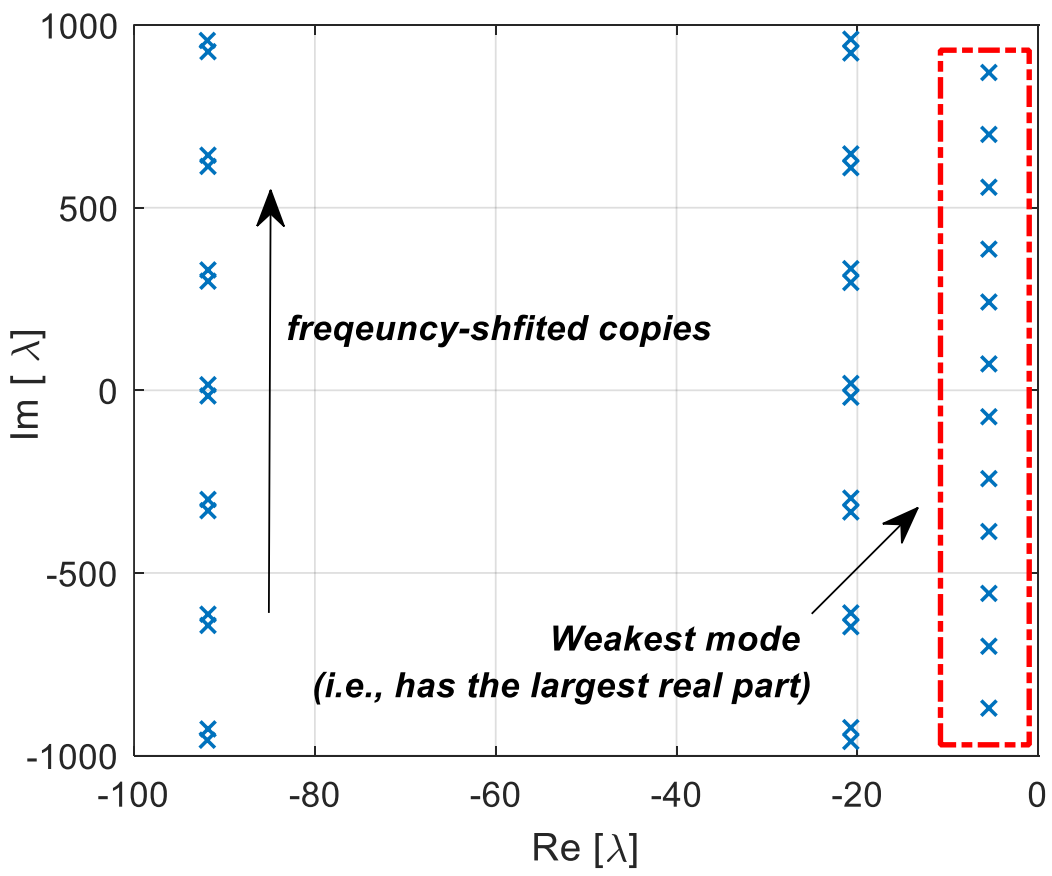

Fig. 5 Eigenvalues of Case system I. Main parameters are given in Table II, where $U_{g\beta} = 0.5\angle{-90^{\circ}}p.u$ is applied.

## D. *Test of Function II: Parametric stability analysis*

Based on the above-presented eigenvalue analysis, impacts of parameter variation on stability can be analyzed by performing a parameter scan against $Re[\lambda_{weakest}]$, this refers to *parametric stability analysis* of this paper. Algorithmically, such studies can be fulfilled by adding outer parameter loops to the AMG solver, as shown in Fig. 6. To better illustrate this, the variation of two parameters will be presented in this work, and the results can be better visualized and characterized as the *parametric stability region* as will be discussed next.

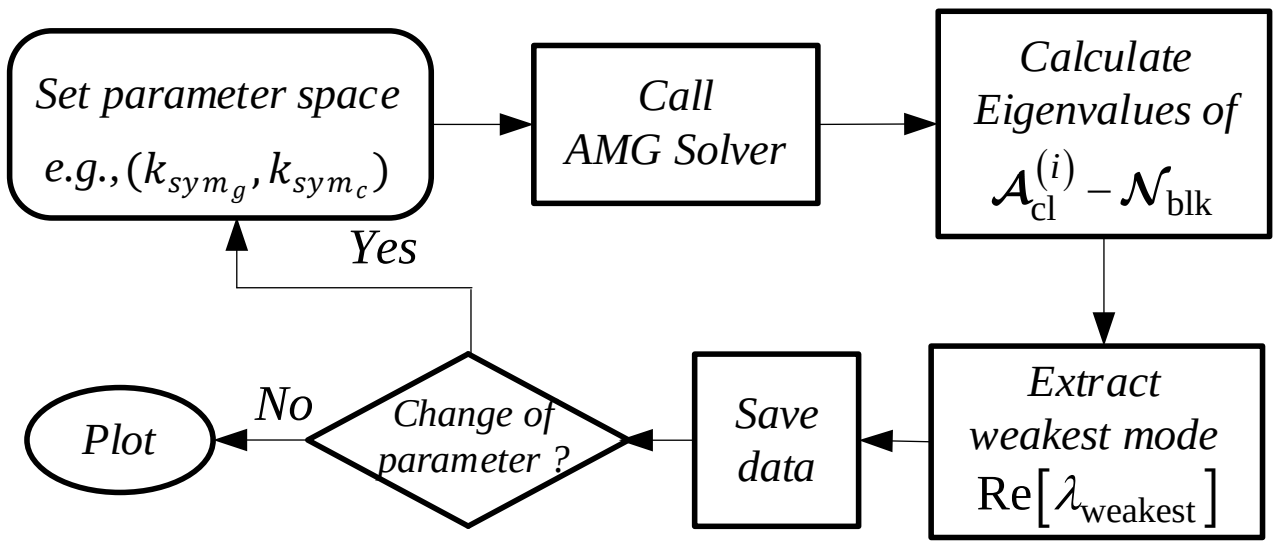

Fig. 6 A flowchart of the parametric stability analysis

### 1) *Stability trait analysis*

In this case, the parameter space $(\alpha_{pll}, |U_{g\beta}|)$ is swept using the algorithm in Fig. 6, while the resulting stability trait plot is shown in Fig. 7 (a). Based on this figure, a qualitative evaluation of the system's stability characteristics can be immediately achieved, e.g., from the figure, one could obtain how will the three-phase asymmetry of grid voltage and the PLL bandwidth affect the closed-loop stability of the grid-VSC system. Furthermore, a quantitative analysis of stability can be achieved by inspecting the values of $Re[\lambda_{weakest}]$ under a given parameter set. For example, $Re[\lambda_{weakest}]$ under the parameter set specified in Fig. 7 (a) is shown to be slightly greater than zero, which means the system will be marginally unstable when operated under this parameter set. This stability conclusion will be verified later.

### 2) *Parametric stability region analysis*

Based on Fig. 7 (a), another form of presentation as mentioned earlier can be obtained, i.e., the parametric stability region. The stability region is more intuitive for analysis that is only concerned with the stability result, i.e., stable or not. It can be obtained by projecting the stability trait plot (e.g., Fig. 7 (a)) onto the plane $z = 0$. In which, projecting the space of $z > 0$ (i.e., $Re[\lambda_{weakest}] > 0$ ) will result in the *unstable* region, otherwise, it will result in the *stable* region.

According to this illustration, the stability trait plot of Fig. 7 (a) can be translated into the stability region plot of Fig. 7 (b), from which the delimitation of the stable and unstable area over the parameter space can be clearly seen, and this boundary implies the critical stability condition of the system. For example, if a particular parameter is found to cross this boundary, it will lead to the change of small-signal stability in the system. This form of presentation can largely assist the parameter designs of a converter when concerned with stability. For instance, according to Fig. 7 (b), if a stable system wants to be assured under various $|U_{g\beta}|$, $\alpha_{pll}$ can be approximately limited below 25 Hz (because the resulting parameter space is contained in the stable region).

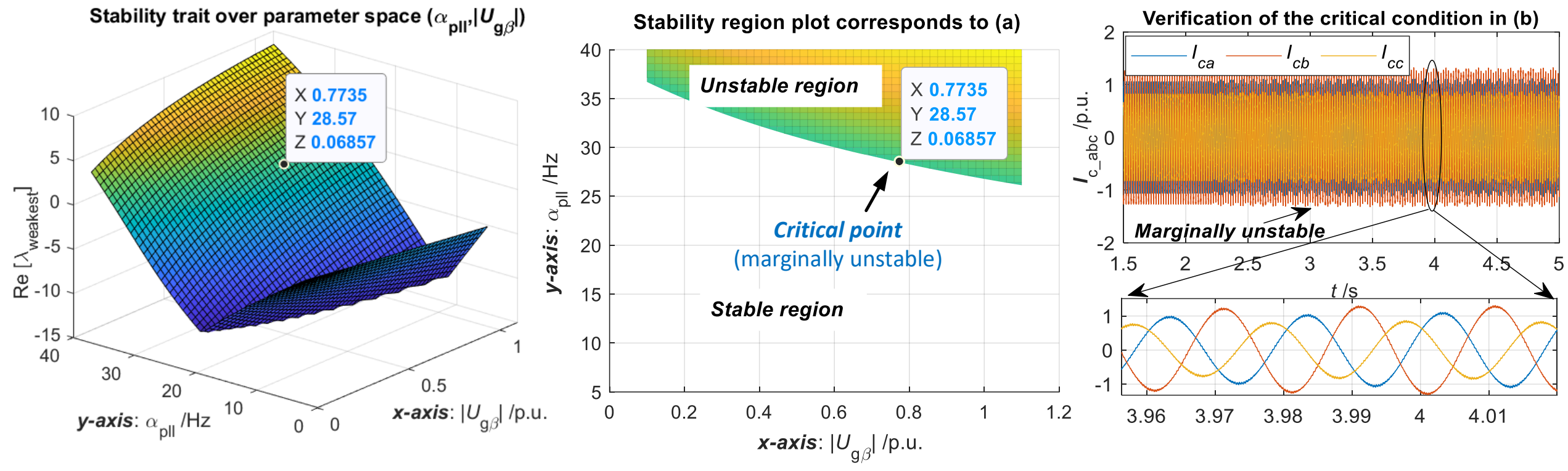

(a) Stability trait plot (b) Stability region plot (c) Time-domain verification of (b)

Fig. 7 Test of parametric stability analysis of Case system I and its time-domain verification in PSCAD/EMTDC (in (a) and (b), $\alpha_c = 200Hz$, $k_{sym_g} = 1$, $k_{sym_c} = 1$, other configurations are given in Table II; in (c), at 2s, parameters are set as the values specified in (b) ).

### 3) *Time-domain verification*

To verify the effectiveness and accuracy of the AMG method in parametric stability analysis, parameters from the theoretically identified critical points in Fig. 7 (b) will be used in simulations to verify whether or not the analysis will accurately predict the stability behavior. The result is shown in Fig. 7(c). According to the current waveforms, it can be seen that the system gradually loses stability when the parameters are set to those under the critical point. This conclusion indicates the theoretic stability analysis of Fig. 7 (b) is correct and thus proves the validity of the AMG method in this analysis.

## E. *Test of Function III: Impedance/admittance generation*

Finally, the frequency response analysis (i.e., Function III of the AMG tool) will be illustrated in this section. To obtain this, the open-loop components characterizing the VSC's small-signal behavior (i.e., $\mathcal{A}_{\text{vsc}}^{(i)},\mathcal{B}_{\text{vsc}}^{(i)},\mathcal{C}_{\text{vsc}}^{(i)},\mathcal{D}_{\text{vsc}}^{(i)}$ ) should be used, which can be obtained along with the PSS extraction via the AMG tool. With these components and by further defining the POC voltage and the VSC current as input and output, the VSC's admittance formulated as the below equation can be obtained, which is essentially a harmonic transfer function (HTF) [17].

$$\mathcal{Y}=\underbrace{\left(\mathcal{C}_{\text{vsc}}^{(i)}\cdot\left(s\mathcal{I}+\mathcal{N}_{\text{blk}}-\mathcal{A}_{\text{vsc}}^{(i)}\right)^{-1}\mathcal{B}_{\text{vsc}}^{(i)}+\mathcal{D}_{\text{vsc}}^{(i)}\right)}_{\mathcal{H}_{\text{vsc}}(s)}\cdot\mathcal{U} \tag{18}$$

if it is written explicitly, the following equations are obtained:

$$\begin{aligned}
&\mathcal{H}_{\text{vsc}}=\begin{bmatrix}\ddots & \vdots & & \vdots & ⋰\\ & \boldsymbol{H}_{(0)}(s_{-1}) & \boldsymbol{H}_{(-1)}(s_0) & \boldsymbol{H}_{(-2)}(s_{+1}) & \\ & \boldsymbol{H}_{(+1)}(s_{-1}) & \boldsymbol{H}_{0}(s_0) & \boldsymbol{H}_{(-1)}(s_{+1}) & \\ & \boldsymbol{H}_{(+2)}(s_{-1}) & \boldsymbol{H}_{(+1)}(s_0) & \boldsymbol{H}_{0}(s_{+1}) & \\ ⋰ & \vdots & & \vdots & \ddots\end{bmatrix}\\
&\mathcal{Y}=\left[\cdots,\boldsymbol{I}_{\text{c}\alpha\beta}(s_{-1}),\boldsymbol{I}_{\text{c}\alpha\beta}(s_0),\boldsymbol{I}_{\text{c}\alpha\beta}(s_{+1}),\cdots\right]^T\\
&\mathcal{U}=\left[\cdots,\boldsymbol{U}_{\text{c}\alpha\beta}(s_{-1}),\boldsymbol{U}_{\text{c}\alpha\beta}(s_0),\boldsymbol{U}_{\text{c}\alpha\beta}(s_{+1}),\cdots\right]^T
\end{aligned} \tag{19}$$

where, $s_k = s + \mathrm{j}k\omega_1$ . Since $\boldsymbol{U}_{\text{c}\alpha\beta}(s_k)=\left[u_{\text{c}\alpha\beta}(s_k),u_{\text{c}\alpha\beta}^{*}(s_k)\right]^T$ and $\boldsymbol{I}_{\text{c}\alpha\beta}(s_k)=\left[i_{\text{c}\alpha\beta}(s_k),i_{\text{c}\alpha\beta}^{*}(s_k)\right]^T$ are 2-by-1 vectors, each entry of $\mathcal{H}_{\text{vsc}}(s)$ is a 2-by-2 matrix, e.g., $\boldsymbol{H}_{(m)}(s_k)$. Moreover, as both the principal-diagonal entries (e.g., $\boldsymbol{H}_{(0)}(s_k)$ ) and the off-diagonal entries (e.g., $\boldsymbol{H}_{(m\neq 0)}(s_k)$ ) in $\mathcal{H}_{\text{vsc}}(s)$ are frequency-shifted transfer functions, it would be sufficient to analyze the components $\boldsymbol{H}_{(0,\pm 1,\ldots \pm m)}(s)$. Next, the frequency responses of $\boldsymbol{H}_{(0,\pm 1,\ldots \pm m)}(\mathrm{j}\omega), m=4$ , under both symmetric and asymmetric three-phase conditions are plotted in Fig. 8.

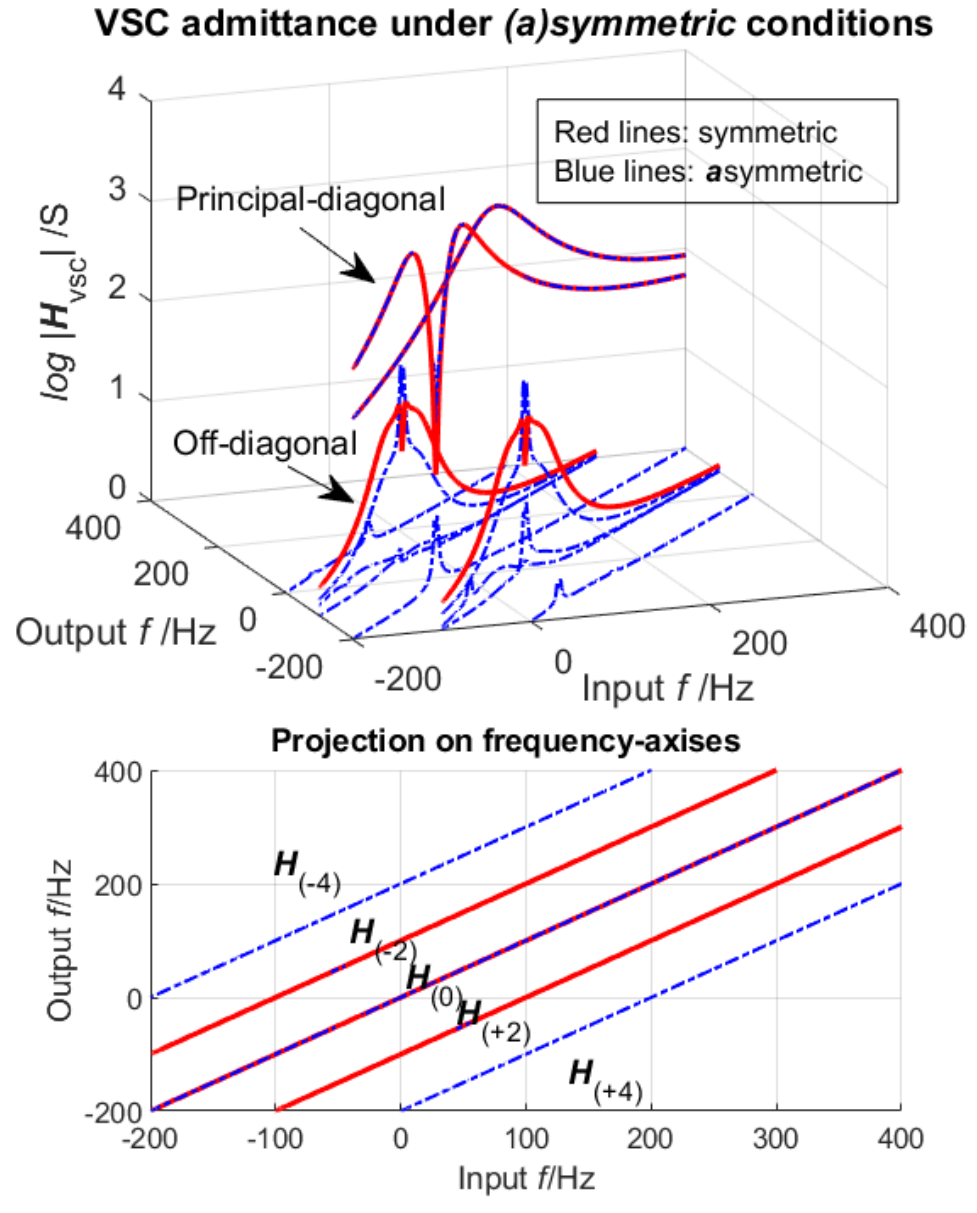

Fig. 8 Test of the frequency response analysis ($|U_{g\beta}| = 0.1\ p.u$ is applied for the asymmetric case, other parameters are given in Table II)

From the plot, first, the multi-frequency coupling behavior of the VSC can be clearly seen (see the projected figure, one input frequency can result in outputs of multiple frequencies), this is another typical trait of the LTP system. Also, it can be noticed that more components exist if the VSC is operated under three-phase asymmetric conditions. In which, the principal-diagonal under the symmetric condition only contains two responses (see the red-lines), while one more response at the bottom emerges with three-phase asymmetries (see the blue-line). Besides, for the symmetric case, the responses of the off-diagonal are away from the principal-diagonal by 100 Hz, such a behavior can be interpreted as the mirror-frequency coupling effect [11] arising from the asymmetries of $dq$ controls [33] (e.g., the PLL's effect).

The analysis demonstrates that the AMG tool is also applicable for impedance and impedance-based analysis.

## IV. Application of the AMG Method in Higher-order Systems: Example of a More Detailed Grid-VSC System

In this section, a more detailed grid-VSC system (i.e., the "Case system II"), including the $LC$-type filter, the active and reactive power control (i.e., $PQ$ control), and the negative sequence current control will be analyzed to demonstrate the applicability of the AMG tool in higher-order systems' analysis.

### A. Modeling and implementation in AMG tool

#### 1) System description

The overall circuit system and control diagrams is shown in Fig. 9. In which, a dual Second Order Generalized Integrator (SOGI)-based sequence filter [35] is used to extract the positive and negative sequence components of the POC voltage. After this, the PLL can be synchronized to the positive sequence grid

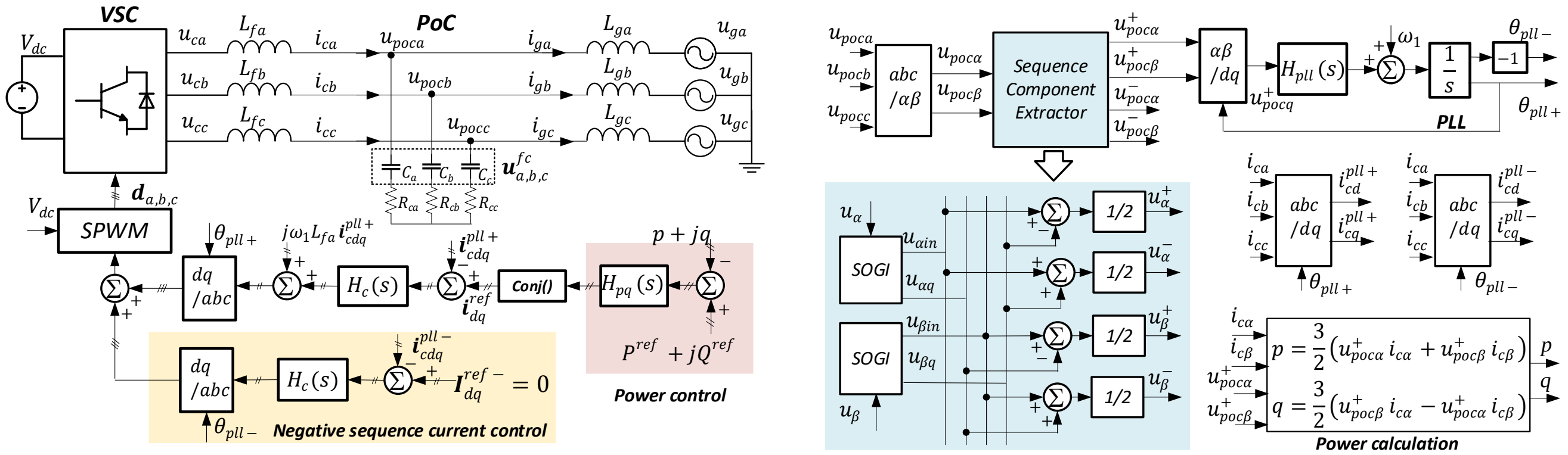

(a) Overall control scheme (b) Other control blocks

Fig. 9 Case system II: the grid-VSC system with unbalanced current control ("$pll\pm$" denote positve and negative rotating dq-frames; "$\pm$"denote positive and negative sequence components)

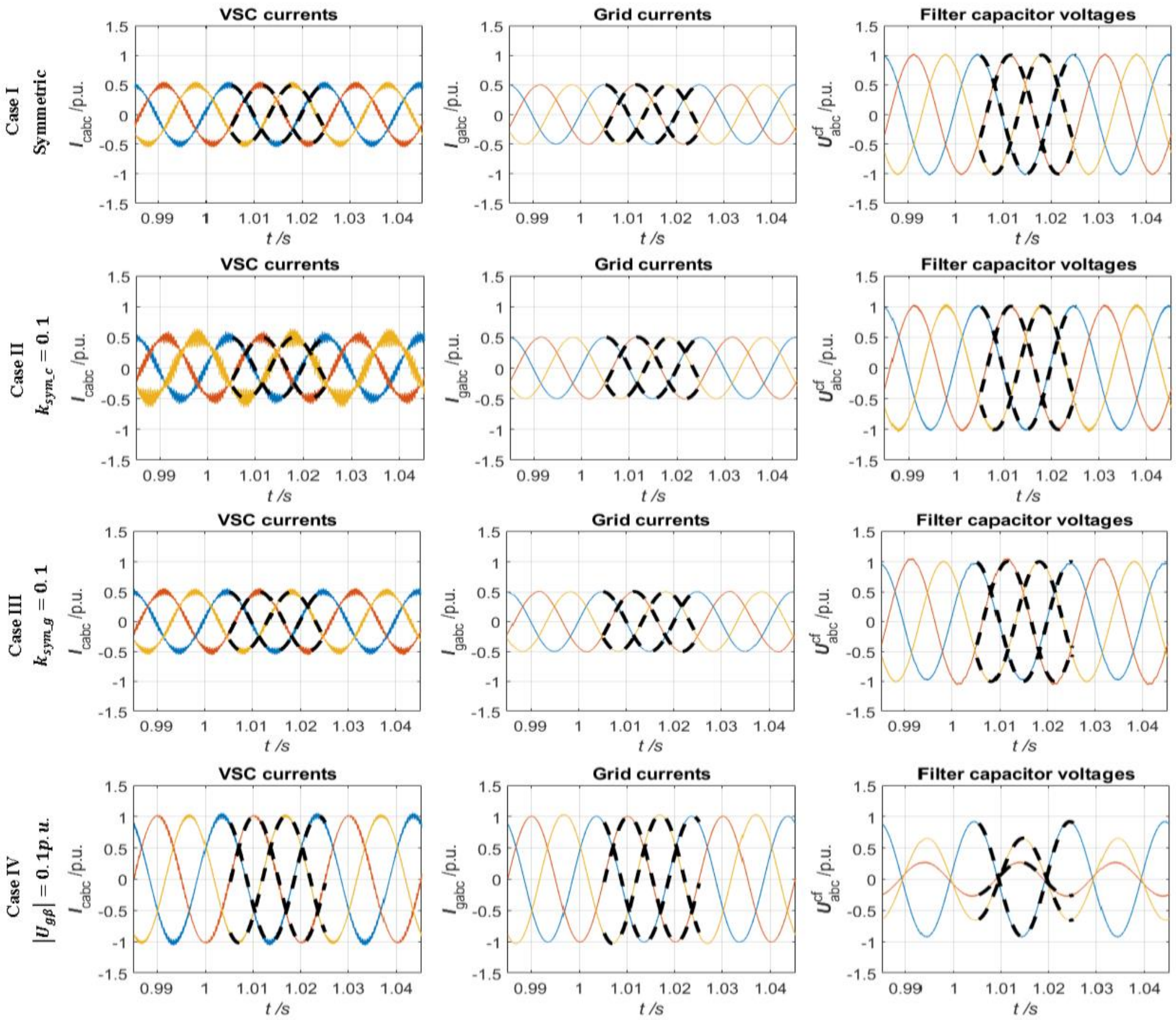

Fig. 10 PSS extraction test for Case system II (dotted lines: AMG, solid lines: PSCAD/EMTDC; $P^{ref}+jQ^{ref}=(0.5+0j)p.u.$; $\alpha_s=20\ Hz$, its relation with the proportional and integral gain is given in appendix.C; $\alpha_c$ for the negative sequence current control is the same as the positive one; the SOGI gain for sequence component extraction is $k_{sogi}=1.414$ ; other circuit and control parameters are listed in Table II)

voltage under unbalanced conditions. Furthermore, the dual-frame current controllers are applied and the negative sequence current of the VSC is controlled to be zero in this study. Hence, the impact of double frequency oscillations (i.e., $2\omega_1$) on the power control feedback can be alleviated by only using the positive sequence voltage for power calculation (see Fig. 9 (b)). Other control schemes for attenuating the $2\omega_1$ oscillation of active or reactive power are documented in [36].

*2) Vector-based state-space model and its implementation in the AMG tool*

Model derivation of Case system II will follow the same procedure as the Case system I. Corresponding vector-based models can be found in Appendix-D. The resulting closed-loop model is in a similar format as (16). However, it should be noted that the model dimension of Case system II is much higher than the Case system I. e.g., the iteration model of Case system I is of order 54 (i.e., $(2N+1)\cdot 6 = 54$, where $N = 4$ is the truncation-order, while the number 6 is the number of state variables); whereas for Case system II, the dimension of the iteration model will become 162 since it has 18 state variables (see Appendix. D).

Then, the closed-loop model of Case system II can be implemented in the AMG tool with a similar procedure given in Section II.E. Some major results will be discussed next.

### *B. Test for PSS extraction*

Running the tool, PSS conditions such as the waveforms of the VSC output currents, the grid currents, as well as the *LC* filter capacitor voltages can be obtained. These waveforms will be compared with those from PSCAD/EMTDC simulations.

First of all, as can be seen from Fig. 10, under all the considered symmetric and asymmetric cases, the AMG results coincide with the simulated waveforms, which confirms the validity of the AMG tool in high-order systems' PSS extraction.

Aside from this main conclusion, it is also noted that owing to the negative sequence current control, the VSC current waveforms can keep balanced under unbalanced system's conditions. Besides, the grid currents are close to the VSC currents but exhibit a better power quality, this is due to the filtering effect of the employed *LC* filter.

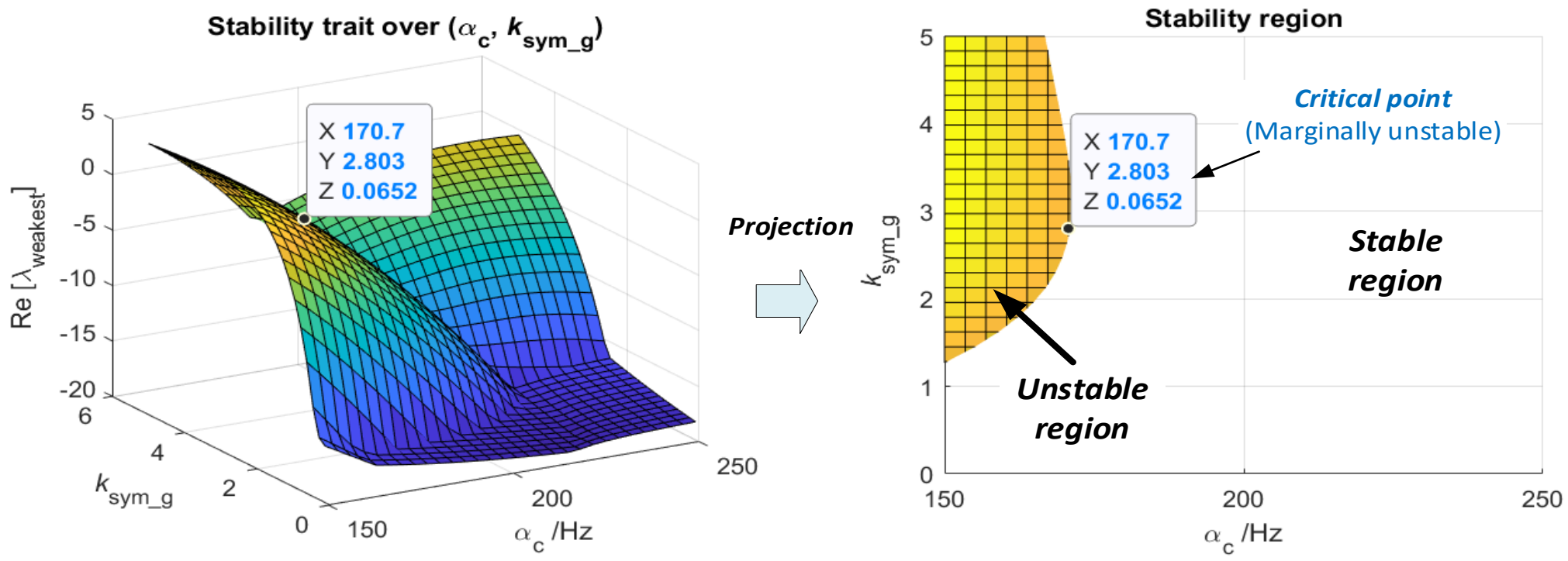

(a) Stability trait and stability region plots. The critical point specified in the plots indicates a marginally unstable system because the value of $Re[\lambda_{weakest}]$ (i.e., the value of *z*-axis) is slightly greater than zero.

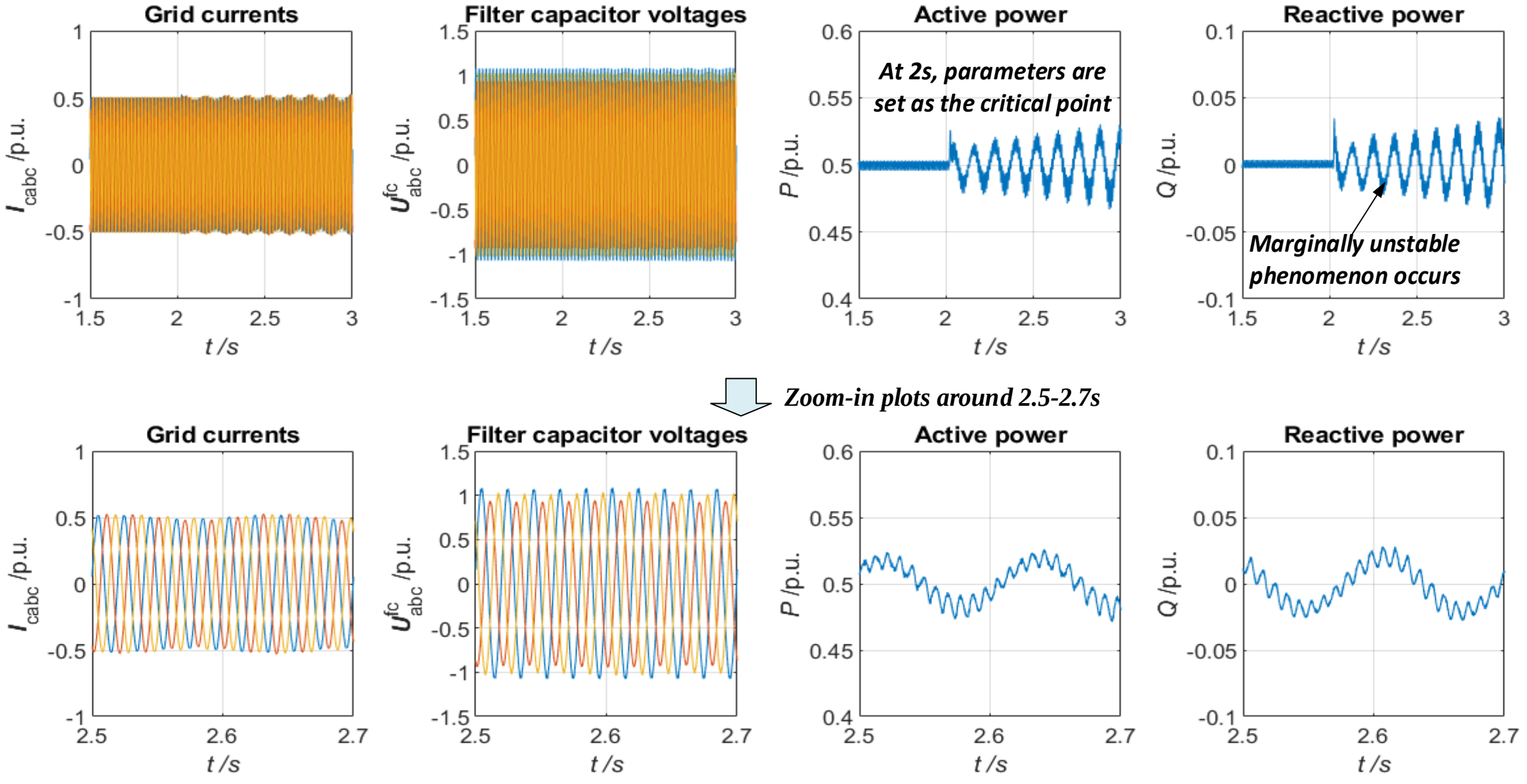

(b) Time domain verification of the above identified critical point

Fig. 11 Parametric stability test for Case system II ($k_{sym_c} = 1$, $|U_{\alpha\beta}| = 1\ p.u.$; $P^{ref} + jQ^{ref} = (0.5 + 0j)\ p.u.$; $\alpha_s = \alpha_{pll} = 20\ Hz$; $k_{sogi} = 1.414$; other parameters are listed in in Table II; generation of these results can be fulfilled by the algorithm shown in Fig. 6)

### C. Test for parametric stability analysis

The parametric stability analysis of Case system II along with its time-domain verification is presented in Fig. 11, where the parameter space $(\alpha_c, k_{sym_g})$ is adopted in this study.

First, according to the stability trait plot given in Fig. 11 (a), it is easy to acquire an overview of the stability characteristics of Case system II under various $\alpha_c$ and $k_{sym_g}$. From this starting point, more detailed analysis of the stability characteristic can be performed. For instance, the plot shows how an increase in $k_{sym_g}$ under a relatively small $\alpha_c$ (e.g., around 150 Hz) will lead to an increase in $Re[\lambda_{weakest}]$, implying that the stability margin will be reduced.

Further, from the stability region plot, it can be seen that the *unstable* region of this case study is located at the upper left corner. In this area, the values of $\alpha_c$ are in general small while the values of $k_{sym_g}$ are large. This observation regarding the location of the unstable region in the studied parameter space can be explained by the *weak grid effect* as follows: a large $k_{sym_g}$ implies a large average value of the three-phase impedances, which is one of the typical indications of a weak grid. Thus, the plot also confirms how the VSC is more prone to instability when operated under such weak grid conditions.

To further verify the accuracy of the AMG-based stability analysis, the critical point specified in the stability region plot (i.e., $(\alpha_c = 170,\ k_{sym_g} = 2.8)$) will be verified by simulation. At this point, since $Re[\lambda_{weakest}] = 0.065$ is slightly above zero, it predicts a marginally unstable system. Then, according to the simulation results in Fig. 11 (b), it can be seen that when the parameters are set to those values under the critical point, the system indeed starts oscillating with slowly increased amplitudes, which agrees with the AMG-based prediction.

On the other hand, since the frequency response analysis of Case system II can be fulfilled in a similar way as Case system I (i.e., by manipulating matrices $\boldsymbol{\mathcal{A}}_{\text{vsc}}^{(i)}, \boldsymbol{\mathcal{B}}_{\text{vsc}}^{(i)}, \boldsymbol{\mathcal{C}}_{\text{vsc}}^{(i)}, \boldsymbol{\mathcal{D}}_{\text{vsc}}^{(i)}$), thus corresponding analysis is omitted here.

## V. Conclusions

Proper and fast evaluation of PSS conditions is of crucial importance for parametric stability analysis of LTP systems. To this end, this paper presented a computational framework (i.e., the AMG tool) for automated PSS acquisition, small-signal model generation and various LTP analyses (e.g., parametric stability analysis) of VSCs. Main conclusions are:

1) The PSS conditions can be efficiently extracted by the developed frequency-domain iteration process. In addition, based on the by-products of this process, a series of LTP analyses, e.g., eigenvalue analysis, parametric stability analysis, and harmonic-domain impedance analysis can be fulfilled, which are demonstrated.

2) The AMG tool simply requires the continuous-domain state-space model of a given system as the input (for VSCs, it refers to the switching-averaged model), while the formulation of the LTP model and the proceeding of the above-listed LTP analyses are automated by the tool. Thus, minimum knowledge about the modeling of LTP systems is required from users.

3) Due to the simplicity in use, it can be applied to other types of converters, e.g., the single-phase converter, the MMC, etc. In which, applications to the stability analysis of a single-phase VSC have been discussed in [37] and [38].

4) Other practical analyses, e.g., the oscillation diagnostics and the stability-oriented control parameter designs, although have not been discussed in detail, can be readily fulfilled according to the presented application examples.

5) At last, although the AMG tool in this paper is realized in MATLAB, it is feasible for being implemented in other suitable software environments.

## Acknowledgement

The authors would like to thank Mr. Sjur Føyen, Mr. Haoxiang Zong, and Mr. Yu Zhang for valuable discussions on this topic and during the course of this work.

## Appendix

### A. Generalized averaging technique

The generalized averaging technique [31] states that any real-valued signal can be arbitrarily approximated by a set of Fourier series with a defined T-period as [31]

$$x(t-T+\tau)=\sum_k \langle x\rangle_k(t)e^{jk\omega_1(t-T+\tau)} \tag{A.1}$$

where the time-varying Fourier coeffect of *k*-th harmonic is

$$\langle x\rangle_k(t)=\frac{1}{T}\int_0^T x(t-T+\tau)e^{-jk\omega_1(t-T+\tau)}d\tau \tag{A.2}$$

where operator $\langle\bullet\rangle_k$ denotes the average of a function against the *k*-th Fourier exponent. The fascinating part of this operator is that it can bring the differential property of time-varying Fourier coefficients:

$$\frac{d\langle x\rangle_k}{dt}=-\mathrm{j}k\omega_1\langle x\rangle_k+\left\langle\frac{dx}{dt}\right\rangle_k \tag{A.3}$$

### B. Transformation matrices

$$\boldsymbol{T}_{\text{abc}/\alpha\beta}=\frac{2}{3}\begin{bmatrix}1 & -\frac{1}{2} & \frac{1}{2}\\ 0 & \frac{\sqrt{3}}{2} & -\frac{\sqrt{3}}{2}\end{bmatrix},\boldsymbol{T}_{\text{abc}/\alpha\beta}=\begin{bmatrix}1 & -\frac{1}{2} & \frac{1}{2}\\ 0 & \frac{\sqrt{3}}{2} & -\frac{\sqrt{3}}{2}\end{bmatrix}^T \tag{A.4}$$

$$\boldsymbol{T}_{\alpha\beta/\text{C}_{\alpha\beta}}=\begin{bmatrix}1 & \mathrm{j}\\ 1 & -\mathrm{j}\end{bmatrix},\boldsymbol{T}_{\text{C}_{\alpha\beta}/\alpha\beta}=\boldsymbol{T}_{\alpha\beta/\text{C}_{\alpha\beta}}^{-1} \tag{A.5}$$

### C. Controller bandwidth and its definition

In this paper, the bandwidth in relation to PI controller gains is defined as: 1) for CC: $k_{\text{pc}}=2\alpha_{\text{c}}L_{fa}, k_{\text{ic}}=2\alpha_c^2L_{fa}$ ; 2) for PLL: $k_{\text{ppll}}=2\alpha_{\text{c}}/U_N, k_{\text{ipll}}=2\alpha_c^2/U_N$ . 3) for the PQ control: $k_{\text{ps}}=\frac{\alpha_{\text{s}}}{1.5U_N\alpha_{\text{c}}}, k_{\text{is}}=\frac{\alpha_{\text{s}}}{1.5U_N}$. Given this arrangement, $\alpha_{\text{c}}$ or $\alpha_{\text{s}}$ is approximately reciprocal to its time constant.

### D. Vector-based state-space model of Case system II

First, the state-vector of Case system II is defined as

$$\boldsymbol{x}_{\mathrm{cl}} = \begin{bmatrix} \vec{\boldsymbol{i}}_{\mathrm{c_\alpha\beta}}, \vec{\boldsymbol{i}}^*_{\mathrm{c_\alpha\beta}}, \vec{\boldsymbol{x}}_{\mathrm{cdq_p}}, \vec{\boldsymbol{x}}^*_{\mathrm{cdq_p}}, \vec{\boldsymbol{x}}_{\mathrm{cdq_n}}, \vec{\boldsymbol{x}}^*_{\mathrm{cdq_n}}, \vec{\boldsymbol{u}}^{\mathrm{fc}}_{\mathrm{\alpha\beta}}, \vec{\boldsymbol{u}}^{\mathrm{fc}*}_{\mathrm{\alpha\beta}}, \\ \vec{\boldsymbol{i}}_{\mathrm{g_\alpha\beta}}, \vec{\boldsymbol{i}}^*_{\mathrm{g_\alpha\beta}}, \vec{\boldsymbol{x}}_{\mathrm{sogi_\alpha\beta}}, \vec{\boldsymbol{x}}^*_{\mathrm{sogi_\alpha\beta}}, \vec{\boldsymbol{x}}_{\mathrm{sogi_q\alpha\beta}}, \vec{\boldsymbol{x}}^*_{\mathrm{sogi_q\alpha\beta}}, \\ x_{\mathrm{pll}}, \delta_{\mathrm{pll}}, \vec{\boldsymbol{x}}_{\mathrm{s_dq}}, \vec{\boldsymbol{x}}^*_{\mathrm{s_dq}} \end{bmatrix} \tag{A.6}$$

where the notation $\vec{\boldsymbol{i}}_{\mathrm{c_\alpha\beta}} = i_{\mathrm{c_\alpha}} + \mathrm{j} i_{\mathrm{c_\beta}}$ denotes the vector representation of the VSC's output current and $\vec{\boldsymbol{i}}^*_{\mathrm{c_\alpha\beta}} = i_{\mathrm{c_\alpha}} - \mathrm{j} i_{\mathrm{c_\beta}}$ denotes its conjugate. This definition applies to other state variables, e.g., $\vec{\boldsymbol{x}}_{\mathrm{cdq_p}}, \vec{\boldsymbol{x}}^*_{\mathrm{cdq_p}}$ are the states of current controllers; $\vec{\boldsymbol{u}}^{\mathrm{fc}}_{\mathrm{\alpha\beta}}, \vec{\boldsymbol{u}}^{\mathrm{fc}*}_{\mathrm{\alpha\beta}}$ are states of the ac capacitor filter voltage; $\vec{\boldsymbol{i}}_{\mathrm{g_\alpha\beta}}, \vec{\boldsymbol{i}}^*_{\mathrm{g_\alpha\beta}}$ are states of the grid current; $\vec{\boldsymbol{x}}_{\mathrm{sogi_\alpha\beta}}$ and $\vec{\boldsymbol{x}}_{\mathrm{sogi_q\alpha\beta}}$ are the states of the SOGI in-phase and quadrature integrators, while $\vec{\boldsymbol{x}}^*_{\mathrm{sogi_\alpha\beta}}, \vec{\boldsymbol{x}}^*_{\mathrm{sogi_q\alpha\beta}}$ are their conjugates; $\vec{\boldsymbol{x}}_{\mathrm{s_dq}}, \vec{\boldsymbol{x}}^*_{\mathrm{s_dq}}$ are the states of *PQ* controller. At last, $x_{\mathrm{pll}}$ and $\delta_{\mathrm{pll}}$ are the two real-valued states of the PLL.

Based on this notation of state variables, next, according to Fig. 9, the state-space models for each component of Case system II can be derived as follows.

*1) The dual SOGI-based sequence extractor*

$$\begin{aligned} \dot{\vec{\boldsymbol{x}}}_{\mathrm{sogi_\alpha\beta}} &= \omega_1 k_{\mathrm{sogi}} \left( \vec{\boldsymbol{u}}_{\mathrm{poc_\alpha\beta}} - \vec{\boldsymbol{x}}_{\mathrm{sogi_\alpha\beta}} \right) - \omega_1^2 \vec{\boldsymbol{x}}_{\mathrm{sogi_q\alpha\beta}} \\ \dot{\vec{\boldsymbol{x}}}_{\mathrm{sogi_q\alpha\beta}} &= \vec{\boldsymbol{x}}_{\mathrm{sogi_\alpha\beta}} \end{aligned} \tag{A.7}$$

Please note that the state equations for $\vec{\boldsymbol{x}}^*_{\mathrm{sogi_\alpha\beta}}$ and $\vec{\boldsymbol{x}}^*_{\mathrm{sogi_q\alpha\beta}}$ are omitted in (A.7). This is because these two equations are simply the conjugates of (A.7) (see such examples in the context of Case system I in Section III. A). Therefore, to avoid the duplication of similar equations and to save some space, the conjugate equations in the following models are all omitted.

The positive and negative sequence voltage at the POC can be extracted (correspond to Fig.10 (b)) as:

$$\begin{aligned} \vec{\boldsymbol{u}}_{\mathrm{poc_\alpha\beta p}} &= \left( \vec{\boldsymbol{x}}_{\mathrm{sogi_\alpha\beta}} + \mathrm{j}\omega_1 \cdot \vec{\boldsymbol{x}}_{\mathrm{sogi_q\alpha\beta}} \right) / 2 \\ \vec{\boldsymbol{u}}_{\mathrm{poc_\alpha\beta n}} &= \left( \vec{\boldsymbol{x}}_{\mathrm{sogi_\alpha\beta}} - \mathrm{j}\omega_1 \vec{\boldsymbol{x}}_{\mathrm{sogi_q\alpha\beta}} \right) / 2 \end{aligned} \tag{A.8}$$

*2) PLL model*

The structure is the same as (10) and (11) the difference is that input of the PLL is the positive sequence voltage vector $\vec{\boldsymbol{u}}_{\mathrm{poc_\alpha\beta p}}$ and its conjugation $\vec{\boldsymbol{u}}^*_{\mathrm{poc_\alpha\beta p}}$ for Case system II.

*3) LC filter model*

The filter capacitor voltage model:

$$\dot{\vec{\boldsymbol{u}}}^{\mathrm{fc}}_{\mathrm{\alpha\beta}} = \left( \vec{\boldsymbol{i}}_{\mathrm{c_\alpha\beta}} - \vec{\boldsymbol{i}}_{\mathrm{g_\alpha\beta}} \right) / C_{\mathrm{f}} \tag{A.9}$$

where the notation $C_a = C_b = C_c = C_{\mathrm{f}}$ is used. The POC voltage can be further written as:

$$\vec{\boldsymbol{u}}_{\mathrm{poc_\alpha\beta}} = \vec{\boldsymbol{u}}^{\mathrm{fc}}_{\mathrm{\alpha\beta}} + R_{\mathrm{cf}} \left( \vec{\boldsymbol{i}}_{\mathrm{c_\alpha\beta}} - \vec{\boldsymbol{i}}_{\mathrm{g_\alpha\beta}} \right) \tag{A.10}$$

where the notation $R_{\mathrm{ca}} = R_{\mathrm{cb}} = R_{\mathrm{cc}} = R_{\mathrm{cf}}$ is used.

*4) Current controller model*

The positive sequence current controller model is:

$$\dot{\vec{\boldsymbol{x}}}_{\mathrm{cdq_p}} = k_{\mathrm{ic}} \left( \vec{\boldsymbol{i}}_{\mathrm{ref_p}} - e^{-\mathrm{j}\delta_{\mathrm{pll}}} e^{-\mathrm{j}\omega_1 t} \vec{\boldsymbol{i}}_{\mathrm{c_\alpha\beta}} \right) \tag{A.11}$$

$$\vec{\boldsymbol{u}}_{\mathrm{c_\alpha\beta p}} = k_{\mathrm{pc}} \left( \vec{\boldsymbol{i}}_{\mathrm{ref_p}} e^{\mathrm{j}\delta_{\mathrm{pll}}} e^{\mathrm{j}\omega_1 t} - \vec{\boldsymbol{i}}_{\mathrm{c_\alpha\beta}} \right) + e^{\mathrm{j}\delta_{\mathrm{pll}}} e^{\mathrm{j}\omega_1 t} \vec{\boldsymbol{x}}_{\mathrm{cdq}} + \mathrm{j}\omega_1 L_{\mathrm{f}} \vec{\boldsymbol{i}}_{\mathrm{c_\alpha\beta}} \tag{A.12}$$

where $\vec{\boldsymbol{i}}_{\mathrm{ref_p}}$ is from the power control that will be shown later.

The negative sequence current controller model is:

$$\dot{\vec{\boldsymbol{x}}}_{\mathrm{cdq_n}} = k_{\mathrm{ic}} \left( \vec{\boldsymbol{i}}_{\mathrm{ref_n}} - e^{\mathrm{j}\delta_{\mathrm{pll}}} e^{\mathrm{j}\omega_1 t} \vec{\boldsymbol{i}}_{\mathrm{c_\alpha\beta}} \right) \tag{A.13}$$

$$\vec{\boldsymbol{u}}_{\mathrm{c\alpha\beta_n}} = k_{\mathrm{pc}} \left( \vec{\boldsymbol{i}}_{\mathrm{ref_n}} e^{-\mathrm{j}\delta_{\mathrm{pll}}} e^{-\mathrm{j}\omega_1 t} - \vec{\boldsymbol{i}}_{\mathrm{c_\alpha\beta}} \right) + e^{-\mathrm{j}\delta_{\mathrm{pll}}} e^{-\mathrm{j}\omega_1 t} \vec{\boldsymbol{x}}_{\mathrm{cdq_n}} \tag{A.14}$$

where $\vec{\boldsymbol{i}}_{\mathrm{ref_n}} = \vec{\boldsymbol{i}}^*_{\mathrm{ref_n}} = 0$ is applied. The VSC output voltage:

$$\vec{\boldsymbol{u}}^{\mathrm{ref}}_{\mathrm{c_\alpha\beta}} = \vec{\boldsymbol{u}}_{\mathrm{c_\alpha\beta p}} + \vec{\boldsymbol{u}}_{\mathrm{c_\alpha\beta n}} \tag{A.15}$$

*5) VSC filter inductor and grid inductor models*

They are the same as (14) and (15).

*6) Power control and current references*

The complex power and power control model are:

$$\vec{\boldsymbol{s}} = \frac{3}{2} \vec{\boldsymbol{u}}_{\mathrm{poc_\alpha\beta p}} \vec{\boldsymbol{i}}^*_{\mathrm{c_\alpha\beta}}, \text{ and } \dot{\vec{\boldsymbol{x}}}_{\mathrm{s_dq}} = k_{\mathrm{is}} \left( \vec{\boldsymbol{S}}_{\mathrm{ref}} - \vec{\boldsymbol{s}} \right) \tag{A.16}$$

The positive sequence current references are:

$$\vec{\boldsymbol{i}}_{\mathrm{ref_p}} = k_{\mathrm{ps}} \left( \vec{\boldsymbol{S}}^*_{\mathrm{ref}} - \vec{\boldsymbol{s}}^* \right) + \vec{\boldsymbol{x}}^*_{\mathrm{s_dq}} \tag{A.17}$$

Assemble all the above equations along with their conjugates will result in the final closed-loop model of Case system II.